\documentclass[a4paper]{spie}  %>>> use this instead for A4 paper
\usepackage[]{graphicx}

\title{The unlikely rise of masking interferometry:
leading the way with 19th century technology. }

\author{Peter G. Tuthill
\skiplinehalf
Sydney Institute for Astronomy, School of Physics, The University of Sydney, NSW 2006, Sydney, Australia \\
}

\begin{document} 
  \maketitle 

%%%%%%%%%%%%%%%%%%%%%%%%%%%%%%%%%%%%%%%%%%%%%%%%%%%%%%%%%%%%% 
\begin{abstract}
The exquisite precision delivered by interferometric techniques
is rapidly being applied to more and more branches of optical astronomy.
One particularly successful strategy to obtain structures at the scale
of the diffraction limit is Aperture Masking Interferometry, which is 
presently experience a golden age with implementations at a host of large
telescopes around the world. This startlingly durable technique, which 
turns 144 years old this year, presently sets the standard
for the recovery of faint companions within a few resolution elements
from the core of a stellar point spread function. This invited review
will give a historical introduction and overview of the modern status 
of the technique, the science being delivered, and prospects for new 
advances and applications.
\end{abstract}

%>>>> Include a list of keywords after the abstract 


%%%%%%%%%%%%%%%%%%%%%%%%%%%%%%%%%%%%%%%%%%%%%%%%%%%%%%%%%%%%%
\section{The dawn of interferometry}
\label{sec:dawn}  % \label{} allows reference to this section

The 19th century witnessed spectacular progress in our understanding
of light. 
For more than 150 years debate had raged over whether light was composed
of particles or waves.
The proponents of the wave theory were given a dramatic boost in 1802
with Young's double-slit experiment.
The showdown between the corpuscular and wave theories of light came
in 1817 with Fresnel's victory over Poisson at the Acad\'{e}mie des Sciences
spectacularly confirming the wave predictions of a bright spot (variously 
called Arago's or Poisson's spot) on axis behind an opaque circular occulter.
Although this victory proved ephemeral and the match was ultimately 
declared a draw with the advent of quantum mechanics a century later, for
the remainder of the 19th century, the wave theory ruled supreme.

% Fresnel and Arago formulate the Fresnel–Arago laws which construct a mathematical
% formalism around Young's suggestion
% that Light was a *transverse wave* and hence has polarization states;

In the first half of the 19th century there were a number of optical devices 
and experiments constructed using interferometeric principles, however almost
all were investigations into the wave nature of light itself, usually
intended to support the case against the corpuscular theory. 
Very few of these could be described using the modern meaning of the word
{\it Interferometer} -- a device intended to make quantitative measurements 
on a physical system by exploiting the wave nature of light.

Perhaps characteristically, the first person to realize the potential for 
measurement of physical properties using light was the great polymath of the
age, Thomas Young. 
Having a primary profession as a medical doctor, Young invented a device around
1813 that he termed an ``Eriometer''\cite{Young1823}, and which was used to make the 
first measurements of the size of red blood cells.
The device worked by measuring the angular extent of the scattering halo created by a 
screen of particles of unknown size, and using his {\it eriometer} Young recorded a diameter
of 7.1\,$\mu$m for red cells (quite close to the modern value of  7.8\,$\mu$m), before 
going on to correctly identify various halos and ``glories'' as seen in nature around 
bright sources of light, such as the moon, as being due to the same basic physics.

On the continent a few years later in 1818, a device which would look remarkably familiar 
to a modern interferometrist was developed by Fran\c{c}ois Arago (who had maintained a
productive collaboration with Fresnel after confirming the wave theory by actually looking
for and finding the counter-intuitive bright spot).
Arago placed two long tubes behind each aperture in a double-slit, and by drying the air
along the path of one of the tubes, was able to observe 1.5 fringes of phase shift over
a 1\,m path giving an exquisitely accurate measurement of the small (0.3\%) difference in
refractive index between humid and dry air. 

It is testament to how far these pioneers were ahead of their time that 30 years elapsed
before the world caught up with these revolutionary ideas, and the power of optical interferometry
was again harnessed for physical science. 
The leading light of the new generation was Fizeau, and he constructed an interferometric device 
in 1851\cite{Fizeau1851} which was to become a benchmark for all that was to follow.
Fizeau's interferometer addressed the leading scientific question of the day: aiming to measure the 
speed of the earth through the luminiferous aether.
In his writings, Fizeau credits the earlier inspirational work of Arago: ``We are indebted to M. Arago 
for a method based upon the phenomena of interference, which is capable of indicating the most 
minute variations in the indexes of refraction of bodies. The experiments of MM. Arago and Fresnel 
upon the difference between the refractions of dry and moist air, have proved the extraordinary 
sensibility of that means of observation.''
A few years later in 1856, Jules Jamin made high precision measurements of refractive index 
using an ``Interferential-Refractor" setup that bears his name\cite{Jamin1856}, and which (after 
several refinements) evolved into the recognizable modern form of the Mach-Zehnder interferometer
before the end of the century.

It is in this setting of the very first few scientific applications for interferometry, with
the actual word ``Interferometer" yet to be coined, that Fizeau suggested in 1868 that
it might be possible to measure the angular diameters of distant stars\cite{Fizeau1868}.
Experimental realization of this idea came 4 years later with Edouard St\'{e}phan's observations
with the Marseilles 80\,cm reflector.  
After another interval of 17 years, Albert Michelson mounted his comprehensive 
experimental campaign delivering the first actual stellar diameters from Mt Wilson, together with 
a description of the mathematical foundations of stellar interferometry.
The history of this era is given with comprehensive clarity in Peter Lawson's historical 
essay\cite{Lawson00} on stellar interferometry, and so I will not attempt to cover it here.

These early observations were made with what we would now call an aperture masking
interferometer, and before finishing this digression into distant history, it is worth
noting a few points. 
Masking was not only the first stellar interferometry, it was among the first few interferometric
experiments of any kind. 
However in some senses, this was a false dawn for the field.
As the more ambitious longer-baseline devices of Michelson and Pease in the 1920's and 1930's 
hit a host of technical problems, it was realized that precision measurements of stellar
diameters were simply beyond the technology of the day\cite{Lawson00}.
However despite the fact that it was 100 years before Fizeau's original vision could be
realized with modern electronic detectors and control systems, the challenge laid down by
the fundamental science of astronomical observation helped to spur the development of interferometry 
in the late 19th and early 20th century, and the ideas went on to underpin methods now found in 
almost every branch of physics.

%%%%%%%%%%%%%%%%%%%%%%%%%%%%%%%%%%%%%%%%%%%%%%%%%%%%%%%%%%%%%
\section{The modern renaissance of masking interferometry}

After a lapse of more than 60 years, masking interferometry was separately and independently
resurrected by two astronomers on opposite sides of the globe.
Interestingly, both shared a common background in radio astronomy where ideas of aperture
synthesis and closure phase were long established.
John Baldwin of Cambridge University was the first to publish interferometric data demonstrating
the recovery of visibility and closure phase signals in the optical using the University of Hawaii 
88-inch telescope\cite{Baldwin86}.
The results from the Australian team, led by Bob Frater, were published only a few months 
later\cite{Frater86} and described experiments at the 3.9\,m AAT telescope some of which entailed
a very large full size plywood mask on the top ring of the telescope which occulted the primary mirror.
These results came in the context of burgeoning interest in high resolution techniques following
the pioneering Narrabri Stellar Intensity Interferometer\cite{NSII67}, and subsequent development
of Speckle Interferometry\cite{speckle70,Weigelt77}, with the first suggestions to implement
optical closure phase\cite{Rogstad68} and pupil masking\cite{RG73,Brown78} all in the early 1970's.
The Cambridge and Sydney groups were soon joined by a third masking team on the 200\,in Hale 
Telescope\cite{Readhead88}, again led by scientists with a strong radio pedigree, producing high 
dynamic range, resolved images of close binary stars\cite{Nakajima89} and evolved giants\cite{Haniff92}.
Around this time there were also variations entailing more complex setups, such as a Dutch sparse-aperture 
shearing interferometer\cite{Bregman88} trialled in La Palma, although few of these progressed past 
prototypes.

Both the Cambridge and Sydney groups continued to actively pursue the development of more advanced 
second-generation masking experiments into the 1990's. 
Following from the first true image recovered from optical aperture synthesis\cite{Haniff87}, 
the Cambridge experiment was moved to the larger 4.2\,m WHT telescope and a novel readout 
mode implemented with the CCD which could deliver a continuous sequence of 1-D snapshot images 
rapid enough to effectively freeze the seeing for bright targets. 
After publishing the first resolved image of a stellar surface (Betelgeuse\cite{Buscher90}), 
the WHT experiment went on to deliver a series of highly productive studies of red giant 
atmospheres and surface structures\cite{Tuthill94,Haniff95,Tuthill97,Wilson97}. 

%-------------
   \begin{figure}
   \begin{center}
   \begin{tabular}{c}
   \includegraphics[height=20cm]{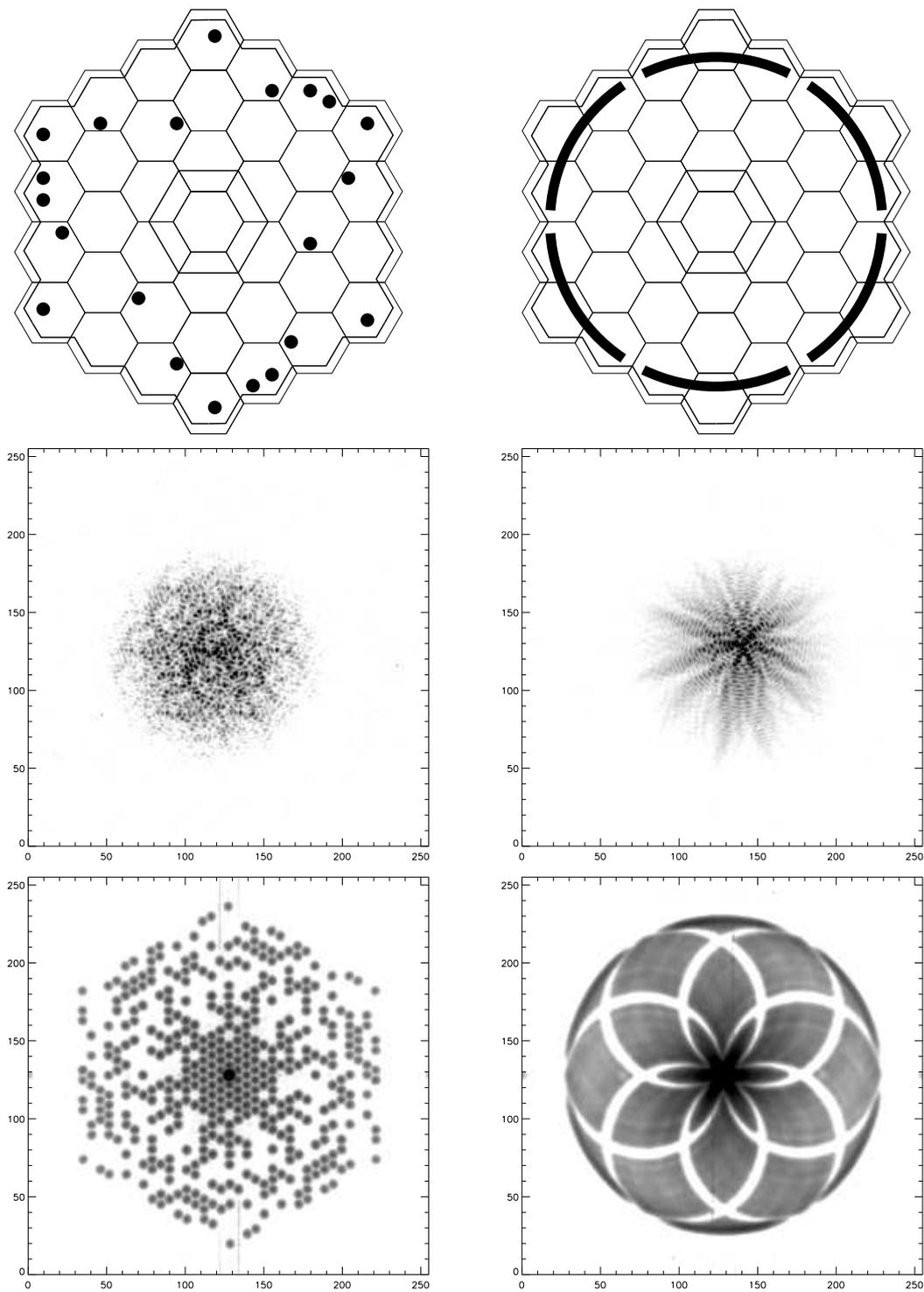}
   \end{tabular}
   \end{center}
   \caption[example] 
%>>>> use \label inside caption to get Fig. number with \ref{}
   { \label{fig:keckmask} 
Upper Panels: the two most productive aperture mask configurations used for the original Keck-1 masking 
experiment, overplotted on an image of the Keck-1 primary mirror.
The left side shows a strictly non-redundant Golay type sampling, while the right hand panel
shows a thin annular ring. The Center panels show a corresponding single-frame interferogram
recorded on the NIRC camera while observing a bright point source reference star. The bottom
panels show the mean power spectra of such data, accumulated over 100 frames.
}
   \end{figure} 
%-------------

Meanwhile, the Sydney group now led by Gordon Robertson took a different tack with the 
construction of their MAPPIT instrument. 
Maintaining a 2-D readout of their detector array, they took the ambitious step of building
a cross-dispersed system in which simultaneous spectral and spatial information could be
recorded\cite{Robertson91,Marson92,Bedding94}.
As for the WHT experiment, the snapshot 1-D Fourier coverage could be rotated on the sky to sample
the {\it uv} plane.
MAPPIT's powerful combination of cross-dispersed fringe data also delivered scientifically valuable
findings for binary stars\cite{Robertson99} and evolved giants\cite{Bedding97,Jacob00,Jacob04,Ireland04}.

The first of the new generation of 10\,m class telescopes, the Keck~1, was fully operational
by the mid 1990's.
Exploiting the factor of 2 increase in baseline, an aperture masking experiment originating
from the University of California at Berkeley was commissioned at Keck in 1995\cite{keckmask00}.
There were several factors beyond the longer baselines that contributed to the success of 
this project, which delivered continuing output of new science for more than a decade.
Near-infrared array cameras were just becoming available with fast readout and enough 
pixels to sample the two-dimensional Fizeau diffraction patterns (yielding snapshot 2-D imaging).
Operation in the near-IR not only exploits the more benign atmospheric seeing, it also
opens a window of key astrophysical relevance.
While the number of stellar targets with apparent diameters resolvable with a mask in the 
optical is only a dozen or so, there are an order of magnitude more systems exhibiting 
hot circumstellar dust which present a large enough angular extent be imaged in the near-IR.
Not only was the experiment well suited to the traditionally-favored interferometric targets
of evolved/dusty red giants\cite{vycma99,raqr00,cit6-00,10216-00,Tuthill05,Woody08,Woody09}, it 
also delivered unique new images of YSOs\cite{lkha01,Bittar01,mwc01,lkha02} and revealed spectacular 
plumes and wakes around colliding-wind binaries\cite{wr10499,wr98a99,WR14002,qt06,wr10408}.
Figure~\ref{fig:keckmask} depicts the two principal workhorse masks delivering this science: 
typically the 21-hole Golay mask was best suited to bright targets (Kmag~$<$~0) while the annulus mask
was suited to fainter objects. 
In addition to these, there was also a 1-dimensional mask which could be used in concert with
a cross-dispersing grism element to deliver simultaneous spatio-spectral\cite{Woody09} data in a 
near-IR analogue of the MAPPIT setup described above.

%%%%%%%%%%%%%%%%%%%%%%%%%%%%%%%%%%%%%%%%%%%%%%%%%%%%%%%%%%%%%
\section{Masking goes global}

Over the last decade, masking interferometry has broadened to encompass wavelength
regimes from the optical to the mid-IR, along the way adopting a host of advanced
modes and improvements to maximize its scientific utility.
Masking projects are now in progress or planned at almost every observatory with 
an 8\,m telescope.

\subsection{Masking and Adaptive Optics}

During the evolution of the experiments described above, of course there were also
many alternate approaches to high angular resolution imaging which were implemented.
One rapidly maturing technology over this interval was Adaptive Optics (AO).
Although at first glance, this may appear to be a direct competitor to the masking interferometry
experiments described, it turned out that the two methods could be combined into a powerful new
form which exploited the strengths of both: {\it sparse aperture adaptive optics}, or perhaps 
more simply, masking-behind-AO.
Among the most persistent outstanding problems in AO has been the robust deconvolution of 
information contained within the of the core of the PSF: a realm within a few $\lambda$/D over
which masking interferometry has been demonstrated to excel due to its emphasis on precision
calibration of residual phase noise.
On the other hand, seeing-limited aperture masking is strictly confined to bright-target science
by the requirement to freeze the atmospheric turbulence in a single exposure, so that 
integration times can be no more than a hundred milliseconds or so. 
Combining masking and AO can be thought of as a Fizeau interferometer in which the AO system 
performs the role of a fast fringe tracker, locking the fringe on every baseline to a phase 
center now defined by the wavefront sensor. 
As long as the AO system is able to deliver some reasonable degree of correction on all
baselines, the fringes are stabilized and one can perform arbitrarily long integrations with 
the science camera, thereby reaching a fainter class of target.
The first experiments to place a mask into a large telescope equipped with an AO system were
implemented on the Hale~200~inch on the PHARO camera in 2003 and the NIRC~2 camera at Keck~2 in
2004\cite{spie06,Lloyd06}, with a further masking experiment commissioned in the NACO instrument 
at the VLT a few years later in 2007 \cite{spie10,Lacour11}.

\subsection{High Contrast detection at High Angular Resolution}

With the faint target reach opened up by sparse-aperture AO, combined with the high-fidelity 
recovery of complex visibility data, masking interferometers could be turned to one of the
key observational problems in contemporary astronomy: resolving high contrast companions 
in the immediate environs of bright stellar targets. 
In particular, it turned out that the self-calibrating properties of the {\it Closure Phases}
-- which could be recovered with errors better than one degree from masking interferometry 
experiments -- was the key to opening a new realm of high contrast science.
Using this powerful observable, unique new science campaigns delivered low-mass stellar binary 
systems and brown dwarfs\cite{Martinache07,coku08,Ireland08,Martinache09}, while the relative
efficiency of observational methods enabled large surveys which pushed the contrasts ever 
higher towards the planetary regime\cite{Kraus08,Bernat10,Kraus11,Hinkley12}. 
Finally, in the last couple of years, compact high contrast detections in very young 
transitional disk systems have been reported \cite{Huelamo11,Kraus12}, the most promising
explanation for which is an actively accreting planetary mass body, possibly attended by 
its own gravitationally entrained shroud or disk of gas and dust\cite{Kraus12}.
It is these high-profile discoveries, perhaps more than anything else, that has motivated
the adoption of masking interferometry so widely as a part of the instrumental capability
modern large telescopes around the world.

\subsection{Mid-Infrared interferometry and segment tilting}

Masks used for high contrast detection, as described in the previous section, tended to have
a relatively small number (7--9) of large ($\sim 1$\,m) holes.
Binary objects are relatively simple and do not require extensive Fourier coverage, while
masks with a few large holes pass more total flux.
Masks of similar design also turn out to be useful for mid-infrared implementations of
masking experiments, although for somewhat different reasons.
In the mid-IR, $r_0$ can be of order several meters in size, and when mask holes are scaled for this
parameter they tend to the few-large-holes side.
A 7-hole mask following these principles was installed in the TRECS camera in Gemini South, and
has been available for masking interferometry since 2007\cite{trecs10}.

With the realization that some large telescope mirrors are comprised of separately steerable
segments, each about the right size to make a workable element in a sparse aperture 
interferometer, the innovative concept of segment tilting interferometry was born.
A great advantage of segment tilting interferometry is that several different subsets of
mirrors can be separately assigned a distinct pointing center on the science camera, so
that the primary mirror is then fragmented into a number of completely independent 
non-redundant interferometer arrays.
Although simple in principle, in practice it turned out to be challenging to maintain
alignment not only of all the separate tilts, but each mirror sub-array also requires
%-------------
   \begin{figure}[htb]
   \begin{center}
   \begin{tabular}{c}
   \includegraphics[height=8cm]{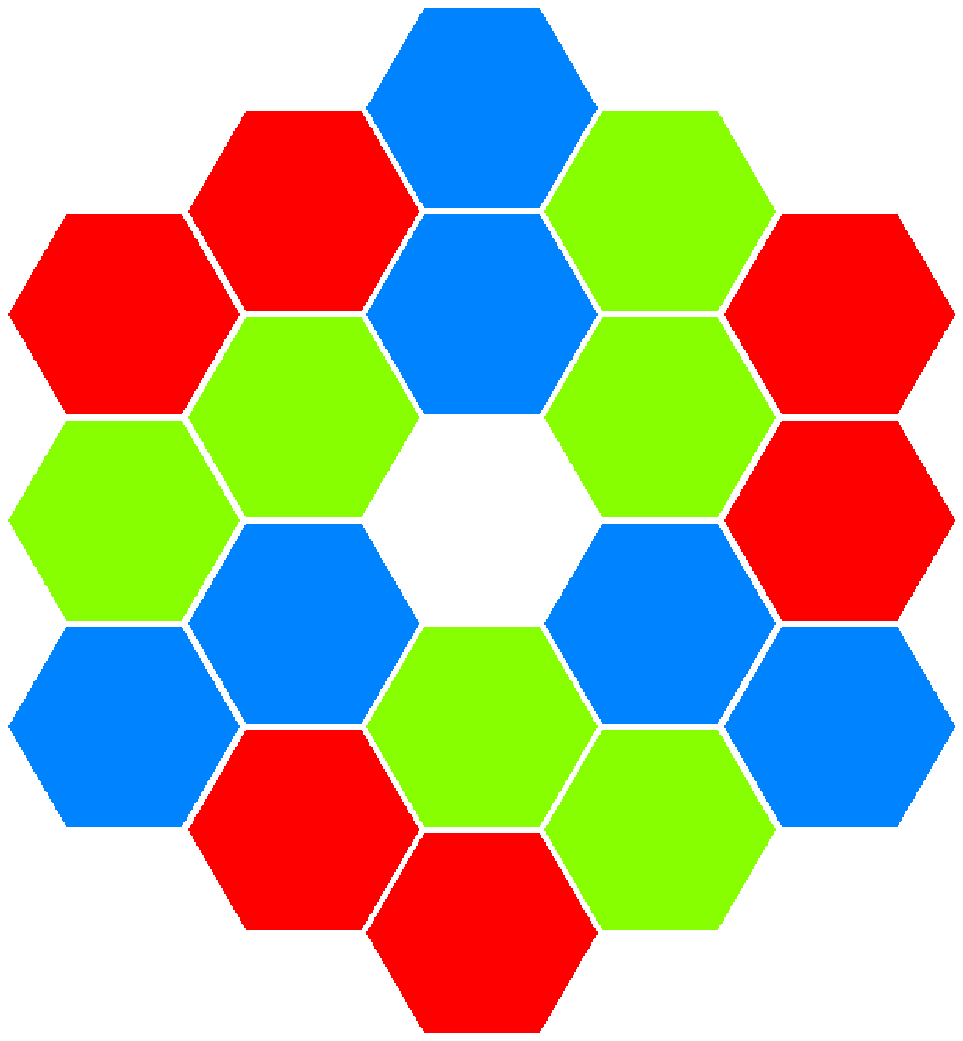}
   \includegraphics[height=8cm]{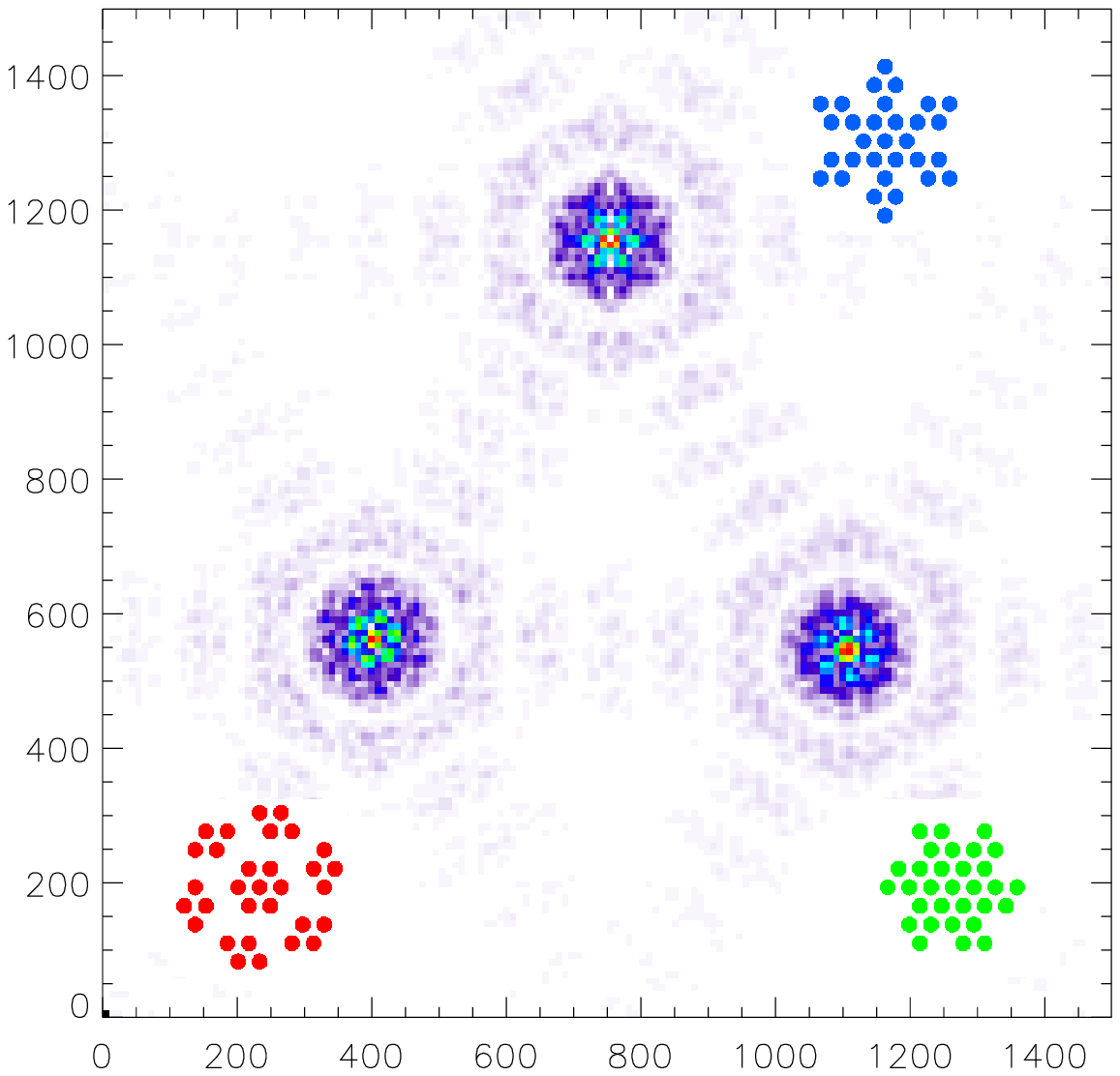}
   \end{tabular}
   \end{center}
   \caption[ex] 
   { \label{jwsttilt} 
Left Panel: A diagram of the 18-segments of the JWST primary mirror, colour coded
so that three separate sub-arrays are formed. Each subset of 6 segments is globally 
tilted about a common point so as to form an image interferogram at a distinct pointing
center on the science detector array, as depicted in the right-hand panel.
Also overplotted here is the representative Fourier coverage for the non-redundant
sampling generated by each subarray.
}
   \end{figure} 
%-------------
specific piston offsets to all segments (all mirrors in a sub-array must lie on
their own distinct parabolic surface).
The first -- and so far only -- segment tilting experiments were performed in 2004/5 at the 
Keck~1 telescope, delivering excellent mid-IR size measurements of YSOs \cite{segtilt09},
however this promising project was cut short with the decommissioning of the LWS instrument
upon which it relied.
However many new-generation large-aperture telescopes, including the JWST as indicated in 
Figure~\ref{jwsttilt}, are being built with segmented apertures making them suited to this
elegant technique.

\subsection{SAMPol: Sparse Aperture Masking Polarimetry}

One advanced masking mode which is presently enjoying growing scientific success is 
the combination of aperture masking with polarimetry enabled by a Wollaston Prism.
In optical setups of this kind, two orthogonal linear polarization states are split 
and recorded simultaneously on the detector array. This results in a situation somewhat
akin to Speckle Holography\cite{Petr98}, where one can record differential observables
between the two separate interferograms with high precision due to the fact that both
beams traversed nearly identical optical paths (including identical instantaneous turbulence) 
before reaching the detector.
This has been exploited to great effect with the NACO instrument\cite{spie10} in recovering 
small differences in visibility arising in orthogonal polarizations due to resolved structures 
in the target.
The power of this unique observational window onto the scattered-light dust distribution
surrounding evolved stars has been demonstrated with the very recent revelation of a
new population of large scattering grains around Mira-type variables\cite{Norris12}, as
described in further detail in the present volume (Norris et al., 8445-2).
To date, these results have exploited only differential visibility amplitudes.
The next step incorporating differential closure phases which should yield even higher 
levels of immunity to noise due to the inherently self-calibrating nature of the
basic observable. 

%  \subsection{Contemporary Masking Facilities}

%%%%%%%%%%%%%%%%%%%%%%%%%%%%%%%%%%%%%%%%%%%%%%%%%%%%%%%%%%%%%
\section{Where Now for Masking Interferometry?}

Although masking interferometry is presently enjoying a small ``Belle \'{E}poque'',
there are many refinements and new generation instruments which will no doubt result
in a continual evolution away from the simplest forms of masking, which would be immediately
recognizable to Fizeau himself, towards more advanced forms.

\subsection{Advanced/Differential Experiments}

Further development of advanced forms of differential masking interferometry is already 
at an advanced stage. Dedicated polarimetric masking experiments, designed for significantly 
greater levels of calibration stability are being considered for deployment within the 
SPHERE/ZIMPOL instrument and also at the Subaru telescope (The VAMPIRES instrument is described
in this proceedings: Norris et al., 8445-2).
These instruments also push masking interferometry back to shorter optical wavelengths where 
polarization signals are generally stronger.

Another area of rapid technological advance has been the implementation of integral field
units (IFU) capable of delivering data cubes which record simultaneous cross-dispersed imaging data. 
The straightforward nature of aperture masking setups means that cost and effort to implement
it, even in the context of a complex IFU setup, is usually minimal.
Recent work detailed in a paper in this meeting (Zimmerman et al., 8445-87) has shown that
exploiting the additional information content in the wavelength direction can deliver a 
factor of a few enhancement in final dynamic range obtained for a masking interferometer. 

\subsection{Developments with AO}

As adaptive optics becomes ever more capable and ubiquitous at a host of the world's
large telescopes, the fundamental nature of problem confronting astronomers wishing to
extract high fidelity imaging data has changed. 
We are rarely now confronted with a completely phase-unstable wavefront, but rather one with
varying degrees of partial long-term coherence. 
This shift will only be amplified with the imminent arrival of extreme-AO systems which
should be deployed at several telescopes worldwide within a year. 
Under conditions of weaker instability of the wavefront phases, alternate approaches to the 
imaging problem may prove themselves competitive or superior to classical aperture masking.
One promising example of this class of algorithms is the {\it Kernel Phase}\cite{Martinache10,Martinache11}
which is also described in this Proceedings (Martinache, 8445-3).

\subsection{The James Webb Space Telescope Interferometer (JWST-I)}

\begin{figure}
   \begin{center}
   \begin{tabular}{c}
   \includegraphics[height=5cm]{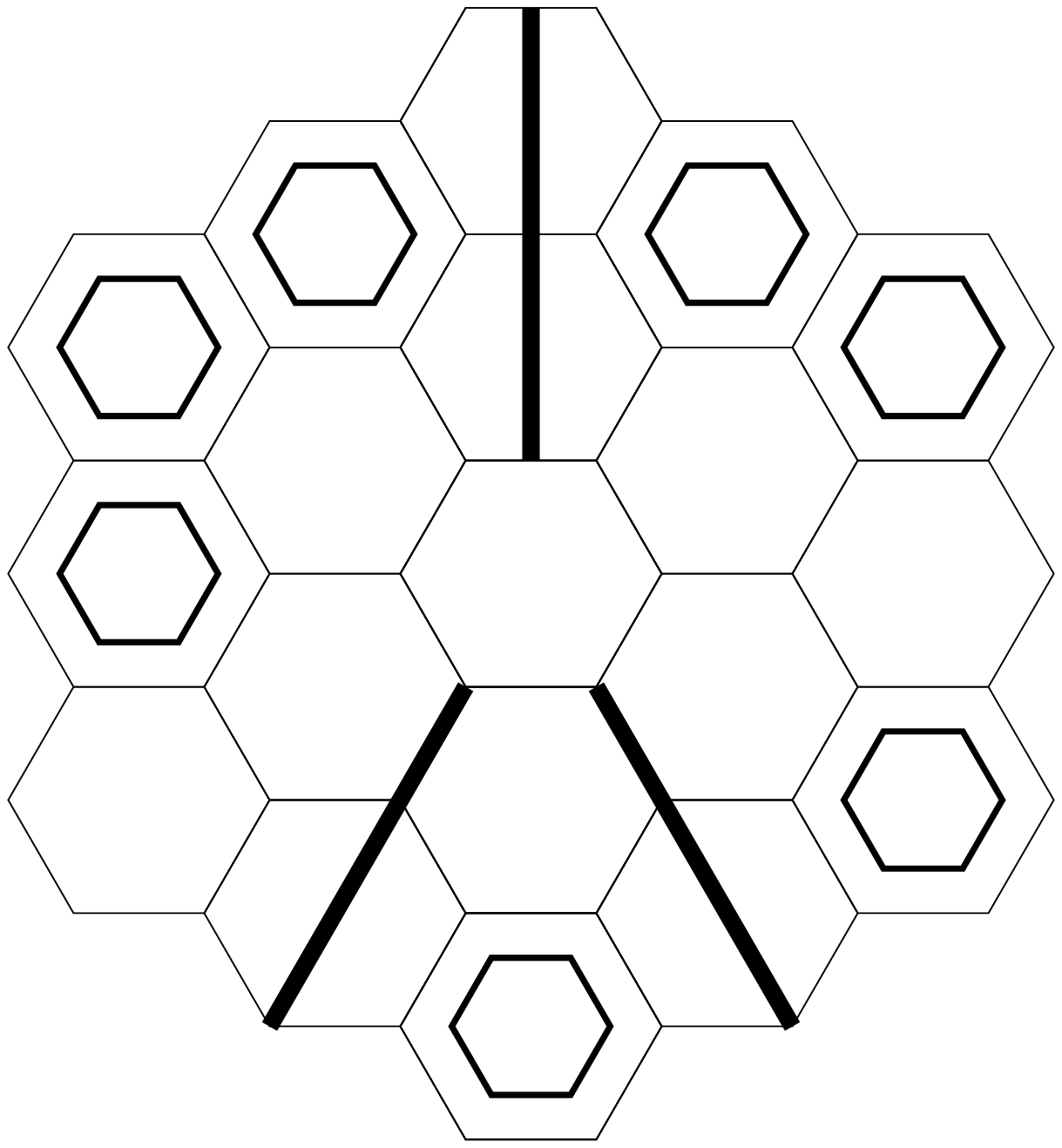}
   \hphantom{XX}
   \includegraphics[height=5cm]{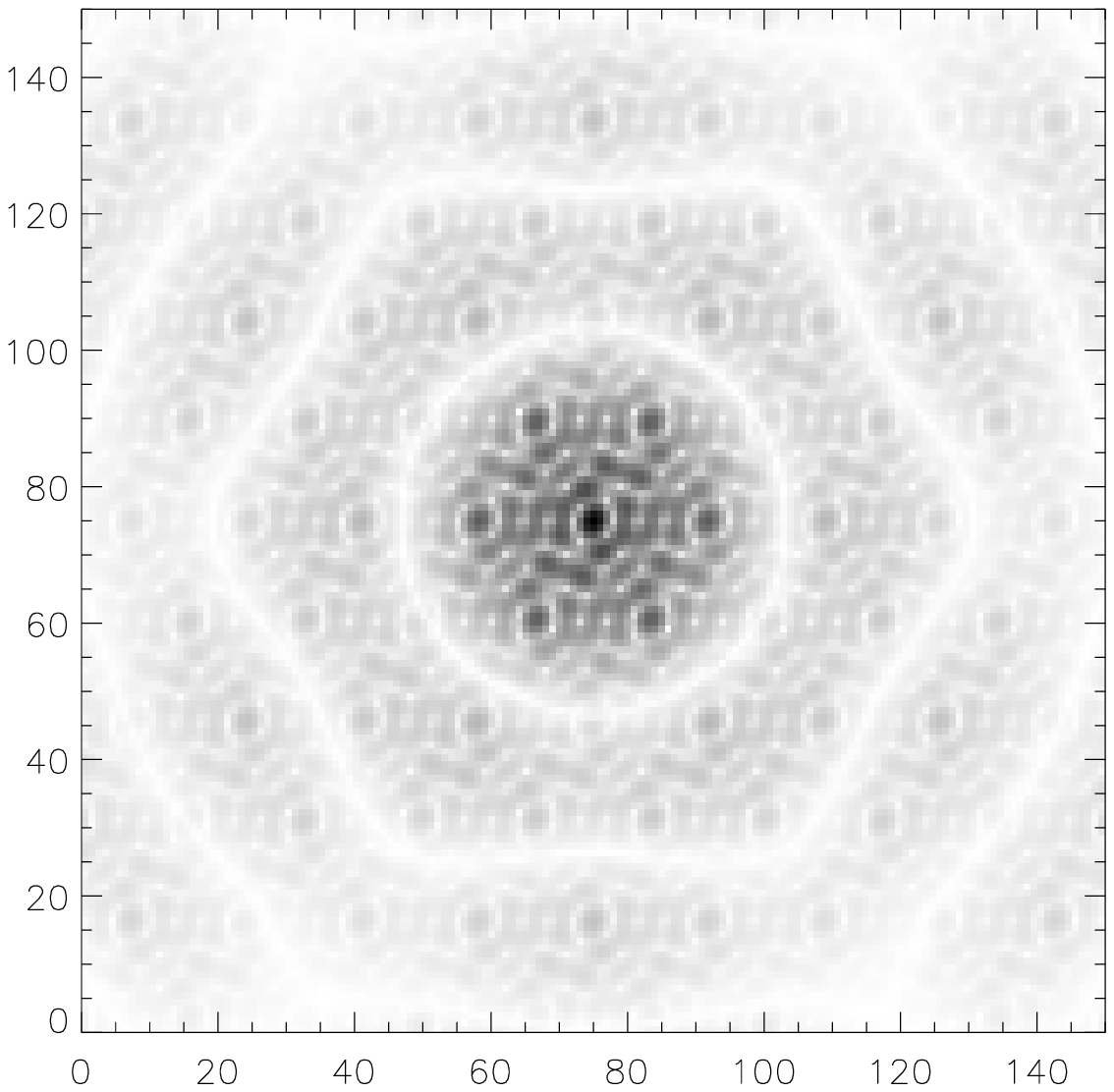}
   \hphantom{XX}
   \includegraphics[height=5cm]{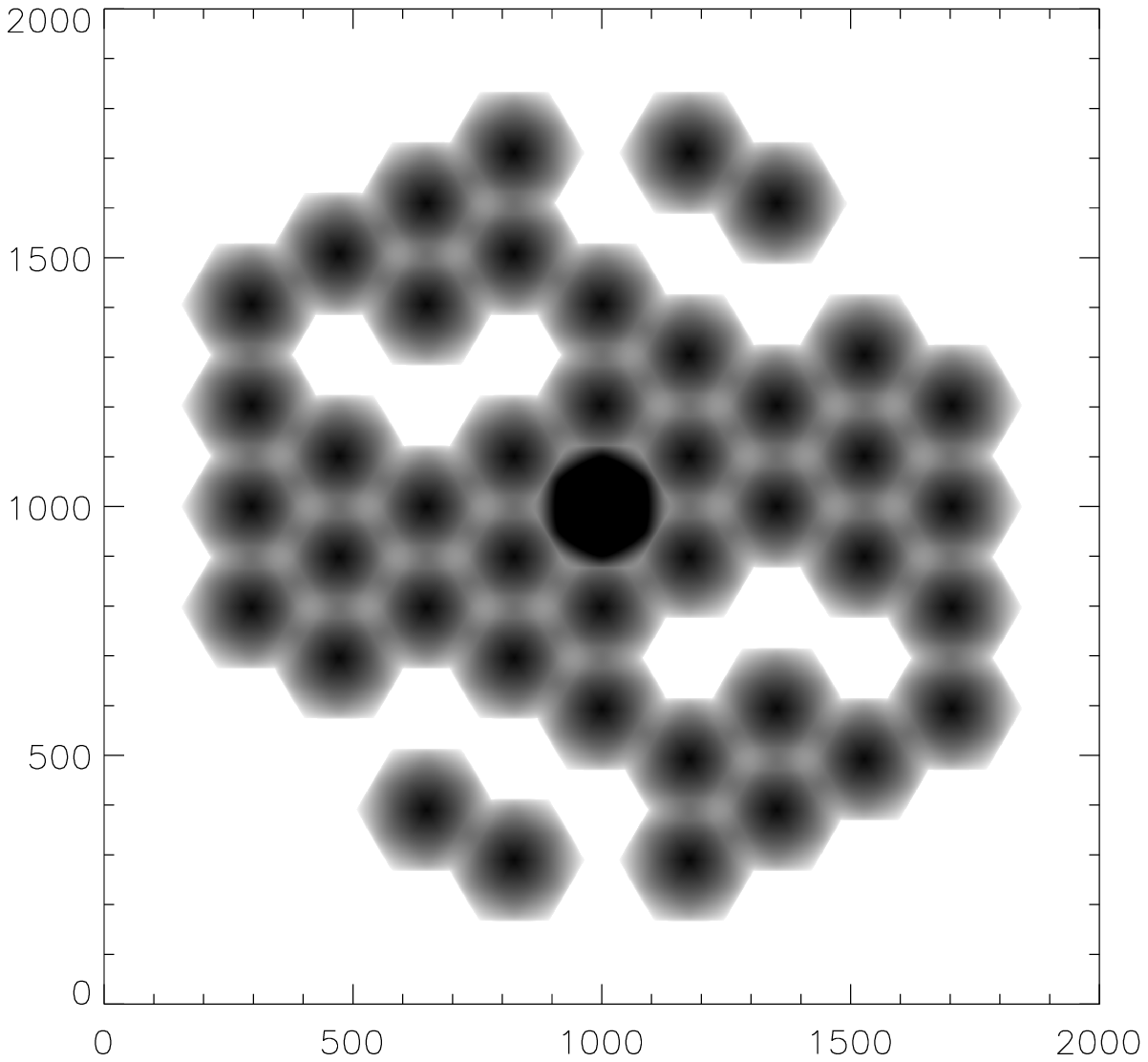}
   \end{tabular}
   \end{center}
   \caption[aa] 
%>>>> use \label inside caption to get Fig. number with \ref{}
   { \label{jwstmask} 
Left Panel: A diagram of the 18-segments of the JWST primary mirror, overlaid with
the locations of the 7-hole mask (hexagonal holes in bold).
Center Panel: A simulated interferogram of a point source star taken with the NIRISS 
instrument (logarithmic stretch).
Right Panel: The power spectrum of the image data, illustrating the non-redundant Fourier
sampling afforded by the aperture mask (logarithmic stretch).
}
\end{figure}

An aperture mask has also been accepted aboard the James Webb Space Telescope as
a part of the NIRISS instrument\cite{jwst10a,jwst10b}.
Although there is obviously no seeing to combat, telescope and instrument
induced speckle noise still dominate the detection threshold for recovery of high contrast companions, 
and simulations show that non-redundant masking can still deliver a significant gain over
conventional imaging methods even in space where wavefront quality and stability are excellent.
A depiction of an aperture mask configuration for JWST is given in Figure~\ref{jwstmask}, and 
further discussion is also given at this meeting by JWST-I lead, Anand Sivaramakrishnan (8442-98).

%-------------

\subsection{Masking Interferometry for Optical Surface Metrology}

The concepts underlying masking's success as a astronomical technique can also find
utility in the context of a solution to the engineering challenge presented by the 
requirement to cophase a segmented primary mirror, such as the JWST. 
A non-redundant pupil, such as is formed by a mask, exhibits the remarkable property
that each Fourier component in the image plane has a direct and unique mapping to a 
pair of subapertures in the telescope pupil.
This key property disentangles the complexity in transforming focal plane images back to 
a knowledge of the pupil from which they arose. 
Using the full telescope pupil, this mapping is an ill-posed problem, but it reduces 
to a few straightforward Fourier measurements when a non-redundant mask is present. 
Simulations which incorporate real-world noise sources indicate that with a few simple
snapshot images, all segments sampled by the mask can be cophased to nanometer precision
in an essentially single-step (not multiply-iterated) process.
The algorithms to accomplish this are described further in 8442-98 (this conference).
Further exploration of this promising technique may also yield a useful method to 
assist with the coming generation of segmented large-aperture ground-based telescopes.

\subsection{Pupil Remapping Interferometry}

Dramatic advances in the control and manipulation of light have arisen from the new
fields of photonics and guided-light optics where technology is strongly driven by 
applications from the communications industry. 
Astronomy has a history of exploiting these advances, and in particular several 
projects around the world have pursued the idea of implementing a non-redundant
beam combiner array by using a photonic remapping of the highly redundant 
sub-pupils. 
Such a setup would reject wavefront phase noise both by way of non-redundancy
and also by spatial filtering of the beams in single-mode waveguide optics;
all the while enabling the full telescope pupil to contribute light to the 
interferometer. 
At least three groups worldwide have been in active development of such a photonic
reformulation of a classical masking interferometer\cite{Perrin06,Lacour07,
Dragonfly10,Mozurkewich11,Huby12}, with papers in the present conference: 8845-4
and 8846-70.
If the considerable technical issues with the stability and throughput of these
challenging devices can be resolved, then they may well inherit the mantle taking
the interferometric data generated by masking interferometer to new levels of
measurement precision.

%%%%%%%%%%%%%%%%%%%%%%%%%%%%%%%%%%%%%%%%%%%%%%%%%%%%%%%%%%%%%
\acknowledgments     %>>>> equivalent to \section*{ACKNOWLEDGMENTS}       
 
I would like to thank the many colleagues who have made the journey in 
high resolution imaging so much fun, and in particular aperture masking 
black belts John Monnier, Sylvestre Lacour, James Lloyd, Chris Haniff, 
Mike Ireland, Frantz Martinache, Barnaby Norris, Gordon Robertson and 
Anand Sivaramakrishnan.

%%%%%%%%%%%%%%%%%%%%%%%%%%%%%%%%%%%%%%%%%%%%%%%%%%%%%%%%%%%%%
%%%%% References %%%%%

\bibliography{Tuthill.bib}   %>>>> bibliography data in report.bib
\bibliographystyle{spiebib}   %>>>> makes bibtex use spiebib.bst

\end{document}